\documentclass[twocolumn]{article}

\usepackage{graphicx}

\title{Diffusion in confined geometries}
\author{P. Sekhar Burada,$^{\rm [a]}$ Peter H\"anggi,$^{* {\rm[a]}}$ Fabio Marchesoni,$^{\rm
  [b]}$\\
  Gerhard~Schmid,$^{\rm [a]}$ and Peter Talkner$^{\rm[a]}$\\[0.2cm]
  $^{\rm [a]}$ Institut f\"ur Physik, Universit\"at Augsburg, \\
  Universit\"atsstr. 1, D-86135 Augsburg (Germany)\\
$^{\rm [b]}$ Dipartimento di Fisica, Universit\`a  di Camerino, \\ 
Via Madonna delle Carceri 9, 
I-62032 Camerino (Italy)\\
$^{\rm *}$  
  Fax: (+49)821 598 3222; 
E-mail: hanggi@physik.uni-augsburg.de\\
}
\date{\today}

\begin{document}

\maketitle

\begin{abstract}
Diffusive transport of particles or, more generally, small objects
is a ubiquitous feature of physical and chemical reaction systems.
In configurations containing confining walls or constrictions
transport is controlled both by the fluctuation statistics of the
jittering objects and the phase space available to their dynamics.
Consequently, the study of transport at the macro- and nanoscales
must address both Brownian motion and entropic effects. With this
survey we report on recent advances in the theoretical and numerical
investigation of stochastic transport occurring either in
micro-sized geometries of varying cross section or in narrow
channels wherein the diffusing particles are hindered from passing
each other (single file diffusion). For particles undergoing biased
diffusion in static suspension media enclosed by confining
geometries, transport exhibits intriguing features such as (i) a
decrease of nonlinear mobility with increasing temperature or, also,
(ii) a broad excess peak of the effective diffusion above the free
diffusion limit. These paradoxical aspects can be understood in
terms of entropic contributions resulting from the restricted
dynamics in phase space. If, in addition, the suspension medium is
subjected to external, time-dependent forcing, rectification or
segregation of the diffusing Brownian particles becomes possible.
Likewise, the diffusion in very narrow, spatially modulated channels
gets modified via contact particle-particle interactions, which
induce anomalous sub-diffusion. The effective sub-diffusion constant
for a driven single file also develops a resonance-like structure as
a function of the confining coupling constant.\\[0.3cm]
\end{abstract}

\section{Introduction}
\label{sec:intro}

Effective control of transport in artificial micro- and
nano-structures requires a deep understanding of the diffusive
mechanisms involving small objects and, in this regard, an operative
measure to gauge the role of fluctuations. Typical such situations
are encountered when studying the transport of particles in
biological cells \cite{Alberts} and in zeolites \cite{zeoliteK},
catalytic reactions occurring on templates or in porous media
\cite{catalytic}, chromatography or, more generally, separation
techniques of size disperse particles on micro- or even nanoscales
\cite{separation}. To study these transport phenomena is in many
respects equivalent to studying geometrically constrained Brownian
dynamics \cite{mazo,chaos05,BM1}. The fact that the diffusion
equation is closely related to the time evolution of the probability
density $P(\vec{x},t)$ to find  a jittering particle at
a location $\vec{x}$ at time $t$ dates back to Einstein's pioneering work on
the molecular-kinetic description of suspended particles
\cite{einstein}.

With this minireview we focus on the problem of the diffusion of
small size particles in confined geometries.  Restricting the volume
of the phase space available to the diffusing particles by means of
confining boundaries or obstacles originates remarkable entropic
effects \cite{liu}. The driven transport of charged particles across
bottlenecks, such as ion transport through artificial nanopores or
artificial ion pumps \cite{siwy1,siwy2,pustovoit,Ilona} or in
biological channels
\cite{berezhkovski,bezrukov,eisenberg1,eisenberg2,eisenberg3} is the
better known model where diffusion is determined by entropic
barriers. Similarly, the operation of artificial Brownian motors
\cite{BM,PT,RH}, molecular motors \cite{leigh} and molecular
machines \cite{balzani} also results from the interplay of diffusion
and binding action by energetic or, more relevant in the present
context, entropic barriers \cite{entropicR}. The efficiency of such
nanodevices crucially depends on the fluctuation characteristics of
the relevant degrees of freedom \cite{machura}. In addition, the
interplay of diffusion over entropic barriers and unbiased time
periodic drives is responsible for certain paradoxical transport
effects, like the recent observation of entropic-diffusion
controlled absolute negative mobility \cite{ANM}.

Although we restricted ourselves to small size particle, we remind
the reader that entropic forces surely affect the dynamics of
extended chains diffusing on a periodic two (2D) or three
dimensional (3D)  substrate. A well established example is
represented by the phonon damping of propagating solitons
\cite{solitons}. Another example of great interest for its potential
applications in nanotechnology is the translocation of a long
polymer molecule through a pore with opening size comparable with
the polymer gyration radius. In this case, entropic effects were
first predicted to theoretically describe the conformation changes
the chain undergoes to move past a conduit constriction
\cite{Arvanitidou} and, then, experimentally observed both in
biological \cite{kasianowicz} and  artificial channels \cite{Han}.

Another instance of constrained Brownian dynamics that rests,
indeed, within the scopes of our minireview, is the so-called single
file diffusion. The motion of an assembly of small size particles in
a narrow channel can be so tightly restricted in the transverse
directions that the particles arrange themselves into a single file.
The longitudinal motion of each particle is thus hindered by the
presence of its neighbors, which act as nonpassing movable
obstacles. As a consequence, interparticle interactions in one
dimension can suppress Brownian diffusion and lead to the emergence
of a new subdiffusive dynamics \cite{SFD-theory}.

The outline of this work is as follows. In Sec. \ref{sec:model} we
detail the geometry of the channels relevant to our review and set
up the mathematical formalism needed to model the diffusion of a
Brownian particle immersed in a confined suspension fluid. In Sec.
\ref{sec:exact} we introduce some exact results for the mobility and
the diffusion coefficient of a driven Brownian particle in a one
dimensional (1D) periodic potential, results that will be handy in
the subsequent sections. In Sec. \ref{sec:4.1} we compute explicitly
the entropic effects on particle transport in a static fluid
filling a periodically modulated channel. In Sec. \ref{sec:4.2} we
address particle transport in a suspension fluid flowing along the
channel subject to stationary pumping. In Sec. \ref{sec:singlefile}
we review numerical and analytical results for the diffusion of a
single file along a periodically corrugated channel both in the
presence and in the absence of an external drive.

\section{Channel models}
\label{sec:model}

\begin{figure}[t]
  \centering
\includegraphics{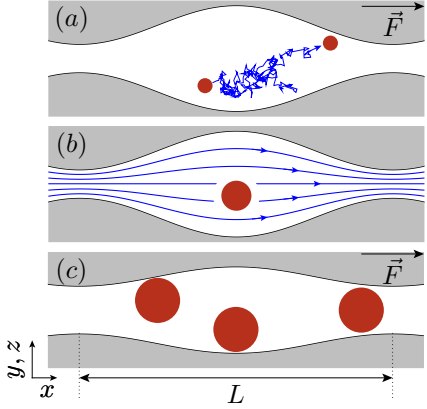}
\caption{(Color online) Brownian particles in a narrow cylindrical channel directed
along the $x$ axis and with periodically modulated boundary $w(x)$
(longitudinal section). (a) Pointlike particle suspended in a static
fluid and subjected to a constant driving force $F$ (Sec.
\ref{sec:4.1}); (b) Single spherical particle dragged along by a
laminar flow (Sec. \ref{sec:4.2}); (c) Single file diffusion of
nonpassing driven particles (Sec. \ref{driven SF}).}
\label{fig:tube}
\end{figure}

We consider the diffusive dynamics of spherical particles in 2D or
3D pores, or channels, which extend in the $x$ direction and possess
a periodically varying cross section. These channels are supposed to
be symmetric with respect to a reflection on the channel axis in 2D
(like those sketched in Fig.~\ref{fig:tube}), or to any rotation
about the channel axis in 3D. The channels are assumed to be
delimited by rigid, smooth walls. The half width of the channel is
described by the well-behaved boundary function $w(x)$. The channel
walls do confine the particles inside the channel but, otherwise, do
not exchange energy with them. In particular, we do not consider the
possibility of adsorption of particles at the walls. Since we will
disregard any rotatory motion of the particles about their centers
of mass, we need not specify particle-wall contact forces. Moreover,
the radius of the particles is supposed to be smaller than the
minima of $w(x)$ (channel bottlenecks), so that the particles are
not restricted to stay within a confined region of the channel, but
rather may diffuse everywhere along the channel.

\subsection{Diffusion equations}
\label{sec:diffequations}

The motion of particles that are immersed in a fluid medium may be
influenced by various types of forces. We will here restrict
ourselves to the discussion of small radius (almost pointlike)
particles, whose presence does not significantly modify the free
motion of the fluid around them (laminar flow regime). The
systematic impact of the fluid on the motion of particles at
position $\vec{x}\equiv(x,y,z)$  is given by the Stokes force
\cite{Lamb}
\begin{equation}
\vec{F}_{\mathrm{Stokes}} = -\gamma [ \dot{\vec{x},t} -
  \vec{v}(\vec{x})]
\label{FStokes}
\end{equation}
where $\vec{v}(\vec{x},t)$ is the instantaneous velocity of the
fluid in absence of the particle, $\dot{\vec{x}}$ is the
instantaneous particle velocity, and
\begin{equation}
\gamma = 6 \pi \eta R
\label{gamma}
\end{equation}
is the friction constant, which is determined by the shear viscosity
$\eta$ of the fluid, and the radius $R$ of the particle. At the same
time the fluid exerts on the particle also a random thermal force
$\vec{F}_{\mathrm{th}}$. In the following, we only deal the case of
fluids with homogeneous temperature $T$ and velocity fields that are
almost constant on the particle scale $R$. As a consequence, to
insure thermalization at temperature $T$ it suffices to set
\begin{equation}
\vec{F}_{\mathrm{th}}({t}) = \sqrt{2 \gamma k_\mathrm{B} T}
\,\vec{\xi}({t})\, , \label{Frand}
\end{equation}
where $k_\mathrm{B}$ denotes the Boltzmann constant and $\vec{\xi}({t})$ is
a standard 3D Gaussian noise with $\langle \vec{\xi}({t}) \rangle =
0$ and $\langle \xi_{i}({t})\,\xi_{j}({t}') \rangle =
\delta_{ij}\,\delta({t} - {t}')$ for $i,j = x,y,z$. Other forces
acting on the particles, like hydrodynamic interactions among
different particles and between single particles and the wall, will
be neglected \cite{cui,6bechinger}. This simplification, in
particular, requires a sufficiently low particle density. 

Finally,
an external force $\vec{F}_{\mathrm{ext}}$ may act on the particles
describing e.g. the gravity force, or in the case of charged
particles, an electrostatic force. We further specialize our
discussion to constant longitudinal external forces pointing in the
direction of the symmetry axis of the channel. The dynamics of the
center of mass $\vec{x}({t})$ of a single particle is then governed
by Newton's equation of motion
\begin{equation}
m \ddot{\vec{x}} = \vec{F}_{\mathrm{ext}} - \gamma
  [\dot{\vec{x}} -\vec{v}(\vec{x},{t})
] + \sqrt{2 \gamma k_\mathrm{B} T} \,\vec{\xi}({t}), \label{eqm}
\end{equation}
where we explicitly allow for a possible time-dependence of the
fluid velocity field. For microparticles moving with typical
velocities of the order of 1 cm/s, the inertial term $m
\ddot{\vec{x}}({t})$ in Eq. (\ref{eqm}) is negligibly small compared
to the environmental forces \cite{Purcell}; therefore, provided that
the fluid velocity does not changes too fast, that is for spectral
frequencies less than a few 100 Hz, one can safely set $m=0$
(overdamped limit or Smoluchowski approximation). Under these
conditions the equation of motion (\ref{eqm}) can be simplified as
\begin{equation}
\dot{\vec{x}} = \vec{v}(\vec{x},{t}) +\frac{1}{\gamma}
\vec{F}_{\mathrm{ext}} +\sqrt{\frac{2
    k_\mathrm{B}T}{\gamma}} \,\vec{\xi}({t}).
\label{LE}
\end{equation}
This Langevin equation is equivalent to the following Fokker-Planck
equation \cite{Risken} for the probability density $P(\vec{x},{t})$
of a particle to be found at the position $\vec{x}$ at time ${t}$:
\begin{equation}
\frac{\partial P(\vec{x},{t})}{\partial {t}} = - \vec{\nabla} \cdot
\vec{J}(\vec{x},t). \label{FPE}
\end{equation}
Here $\vec{J}(\vec{x},t)$ denotes the corresponding probability
current density
\begin{equation}
\vec{J}(\vec{x},{t}) = -\left ( \vec{v}(\vec{x},{t}) +
  \frac{\vec{F}_{\mathrm{ext}}}{\gamma} \right) P(\vec{x},{t}) + \frac{k_\mathrm{B}
  T}{\gamma} \vec{\nabla}  P(\vec{x},{t}).
\label{J}
\end{equation}
These equations have to be supplemented by appropriate boundary
conditions, which will be discussed in the following subsection.

\subsection{Boundary conditions}
\label{sec:bc}

The channel is typically characterized by two boundary regions: A
transverse boundary naturally results from the presence of the
walls, whereas a longitudinal boundary is required to account for
the channel length.

The probability flux normal to the boundary in the presence of a
rigid hard wall must vanish. Thus, to prevent pointlike particles
from leaving the channel or being adsorbed at the walls, we must
impose that
\begin{equation}
\vec{n}(\vec{x}) \cdot \vec{J}(\vec{x},t) = 0 \quad \vec{x} \in
\;\mathrm{wall}, \label{bcw}
\end{equation}
where $\vec{n}(\vec{x})$ denotes the unit vector normal to the wall
at point $\vec{x}$. Note, however, that the center of mass of a
spherical particle of finite radius $R$ may approach the wall only
up to a distance $R$, so that for finite size particles the boundary
conditions (\ref{bcw}) applies on an appropriate inner surface
parallel to the channel walls \cite{Schind}.

Various kinds of boundary conditions exist that regulate the inward
and outward probability flows at the ends of a channel
\cite{Lamb}.
If the channel connects large, well mixed particle reservoirs, then
constant probability densities $P_{L,R}$ may be assigned at the
channel ends. This leads to Dirichlet boundary conditions of the
form
\begin{equation}
P(x_{\mathrm{L}},y,z) = P_{L}, \quad P(x_{\mathrm{R}},y,z) = P_{R},
\label{bceD}
\end{equation}
where $x_{L}$ and $x_{R}$ denote respectively the left and right
endpoints of the channel.
A more detailed description of the particle flow into and out of a
channel can be achieved by relating flux and probability densities
at $x=x_{R,L}$, that is \cite{Bezrukov-radiation}
\begin{eqnarray}
J_{x}(x_{L},y,z,{t}) &=& - \kappa_{L} P(x_{L},y,z,{t})\, , \nonumber \\
J_{x}(x_{R},y,z,{t}) &=& \kappa_{R} P(x_{R},y,z,{t})\, . \label{bcr}
\end{eqnarray}
Here, positive (negative) constants $\kappa_{L,R}$ correspond to
partially absorbing (emitting) boundaries. As special cases,
reflecting and absorbing boundaries correspond to $\kappa_{R,L} = 0$
and $\kappa_{R,L}  = \infty$, respectively.

Finally, for an infinitely long channel the periodic boundary
conditions
\begin{equation}
P(x,y,z,{t}) =  P(x+L,y,z,{t})
\label{bcp}
\end{equation}
are more appropriate \cite{Risken}. In the case of velocity fields
which are constant with respect to time or vary periodically in
time, these boundary conditions allow for stationary flux carrying
solutions \cite{BM,PT,RH}.

\section{Exact results for 1D systems}
\label{sec:exact}

In order to set the stage, we consider first the ideal case where
the diffusion of a particle in a periodically corrugated channel can
be assimilated to the diffusion on an energetic landscape
represented, for simplicity, by a 1D periodic substrate $V(x)$ with
period $L$, namely $V(x+L)=V(x)$. Such a model is often employed to
model, for instance, nanotube \cite{dresselhaus1996} and zeolite
diffusion \cite{zeoliteK}. Let us consider a Brownian particle with
mass $m$, coordinate $x$, and friction coefficient $\gamma$,
subjected to a static external force $F$ and a thermal noise
$F_\mathrm{th}(t)$. In the notation of Sec. \ref{sec:diffequations}, we set
$F_\mathrm{ext}=-V'(x)+F$.

The corresponding stochastic dynamics is described by the Langevin
equation
\begin{equation}
\label{xLE} m\ddot x = - V'(x) - \gamma\,\dot x + F + \sqrt{2\gamma
k_\mathrm{B}T}\,\xi(t),
\end{equation}
where $\xi(t)$ is the standard Gaussian noise also defined in Sec.
\ref{sec:diffequations}. Moreover, for the purposes of this review,
the substrate $V(x)$ can be taken symmetric under reflection,
$V(x)=V(-x)$.

In extremely small systems, particle fluctuations are often
described to a good approximation by the {\it overdamped} limit of
Eq. (\ref{xLE}), i.e., by the massless Langevin equation
\begin{equation}
\gamma \dot x =  - V'(x) + F + \sqrt{2\gamma k_\mathrm{B}T}\,\xi (t)
\, ,
\label{oLE}
\end{equation}
where the inertia term $m\ddot x$ has been dropped altogether
(Smoluchowski approximation).

An overdamped particle is trapped most of the time at a local
minimum of the tilted substrate as long as $F \leq F_3$, $F_3$
denoting the depinning threshold $F_3=\mathrm{max}\{V'(x)\}$. Drift
occurs by rare noise induced hopping events between adjacent minima.
For $F>F_3$ there exist no such minima and the particle runs in the
$F$ direction with average speed approaching the no-substrate limit
$F/\gamma$. This behavior is described quantitatively by the
mobility formula:
\begin{equation}
\label{eq:mobility-def} \mu(F)\equiv \frac{\langle \dot x
\rangle}{F},
\end{equation}
with
\begin{equation}
\label{eq:current-num} \langle \dot x \rangle \equiv
\lim_{t\to\infty}\frac{\langle x(t) \rangle}{t}=\frac{L}{\langle
t(L,F) \rangle}.
\end{equation}
Here and in the following, $\langle t^n(L,F) \rangle$ denotes the
$n$-th moment of the first passage time of the particle across a
substrate unit cell in the $F$ direction.


As the particle drifts subjected to the external force $F$, the
random hops cause a spatial dispersion of the particle around its
average position $\langle x(t) \rangle$. The corresponding {\it
normal} diffusion coefficient,
\begin{equation}
\label{eq:diffusion-num} D(F) \equiv \lim_{t\rightarrow \infty}
\frac {\langle x(t)^2\rangle - \langle x(t)\rangle^2} {2t},
\end{equation}
can be computed analytically by regarding the hopping events in the
overdamped regime as manifestations of a renewal process,
that is \cite{Reimann_PRL}:
\begin{equation}\label{eq:diffusion-def}
D(F)= L^2\,\frac{\langle t^2(L,F)\rangle - \langle t(L,F)\rangle^2}
{2\,{\langle t(L,F)\rangle}^{3}}.
\end{equation}
Simple algebraic manipulations lead to explicit expressions for the
nonlinear mobility \cite{Risken,stratonovich,tikhonov,hanggi},
\begin{equation}
\label{eq:mobility-formula} \mu(F)=
\frac{D_0L}{F}~\frac{1-e^{-LF/k_\mathrm{B}T}}{\int_0^L I_+(x)dx},
\end{equation}
and for the diffusion coefficient \cite{Reimann_PRL},
\begin{equation}
\label{eq:diffusion-formula} \frac{D(F)}{D_0}= L^2~\frac{\int_0^L
I^2_+(x) I_-(x)dx}{[\int_0^L I_+(x)dx]^3}.
\end{equation}
Here, $I_{\pm}(x)=\int_0^L e^{[\pm V(x)\mp V(x\mp y)-yF]/k_\mathrm{B}T}dy$
and $D_0=k_\mathrm{B}T/\gamma$ denotes Einstein's coefficient for a free
diffusing Brownian particle.

For $F\to 0$ Eqs. 
(\ref{eq:mobility-formula}) and (\ref{eq:diffusion-formula}) reproduce the zero-bias identity
$D(0)/D_0=\gamma\mu(0)$ with $\mu(0)=L^2/\left (\gamma \int_0^L
I_+(x)dx \right )$\cite{Risken,Lifson}. Notably, as $F$ approaches
the depinning threshold $F_3$ the mobility curve
(\ref{eq:mobility-formula}) jumps from zero (locked state) up to
close $1/\gamma$ (running state). Correspondingly, the diffusion
coefficient (\ref{eq:diffusion-formula}) develops a diffusion excess
peak, i.e. with $D > D_0$, consistently with numerical observations
in Ref. \cite{Costantini}. Both the mobility step and the $D$ peak
get sharper and sharper as $T$ is lowered \cite{Reimann_PRL}. The
same conclusions apply in the presence of inertia (i.e. for
particles of finite mass $m$) \cite{Costantini,Cattuto}, as well,
with the difference that the relevant depinning threshold shifts
towards lower values (proportionally to $\gamma$ as $\gamma \to 0$
\cite{Risken}).

Finally, we emphasize that the above formulas for the nonlinear
mobility and the effective diffusion coefficient retain their
analytic structure also when generalized to anomalous
(sub)-diffusion on a 1D substrate, by merely substituting the normal
diffusion constant $D_0$ by the fractional diffusion constant
occurring in the corresponding fractional diffusion equation
\cite{goychuk}.

\section{Particle diffusion in a static fluid}
\label{sec:4.1}

In the limiting case of pointlike particles with zero interaction
radius diffusing in constrained 2D or 3D geometries, elastic contact
particle-particle interactions can be neglected. As long as
hydrodynamically mediated interactions can also be neglected, see
e.g. \cite{cui,6bechinger}, the confining action of the channel
walls is modeled by the perfectly reflecting boundary conditions
(\ref{bcw}).

In the presence of a constant external force pointing in the channel
direction, an overdamped Brownian particle suspended in a static
medium is described by the Langevin equation (\ref{LE}), or by the
corresponding Fokker-Planck equation (\ref{FPE}), both with zero
velocity field $\vec{v}(\vec{x},t)\equiv 0$. For a general choice of
the periodic  boundary  $w(x)$, there exist no exact analytical
solutions to the Fokker-Planck equation (\ref{FPE}) with boundary
conditions (\ref{bcw}). However, approximate solutions can be
obtained by reducing the problem of free Brownian diffusion in a 2D
or 3D channel to that of Brownian diffusion on an effective 1D
periodic substrate (Sec. \ref{sec:exact}). Under such a scheme,
narrow channel constrictions, corresponding to geometric hindrances
in the fully dimensional problem, are modeled as entropic 1D
barriers.

In the absence of external forces, i.e., for $\vec{F}_{\rm ext}= 0$,
particle dynamics in confined structures [see
Fig.~\ref{fig:tube}(a)] can be approximately described by the
Fick-Jacobs kinetic equation with spatially dependent diffusion
coefficient \cite{Jacobs, Zwanzig, Reguera_PRE,Percus}:
\begin{equation}
\label{eq:fickjacobs}
\frac{\partial P(x,t)}{\partial t}=\frac{\partial}{\partial
  x}\left(D(x)\,\sigma(x)\frac{\partial}{\partial
    x}\, \frac{P}{\sigma(x)}\right) \, ,
\end{equation}
with $\sigma(x)$ denoting the dimensionless channel cross-section $2
w(x)/L$ in 2D and $\pi w^{2}(x)/L^{2}$ in 3D. Equation
(\ref{eq:fickjacobs}) was obtained \cite{Zwanzig} from the full
Fokker-Planck equation (\ref{FPE}) for small amplitude boundary
modulations $w(x)$, on assuming a uniform density $P(\vec{x},t)$ and
integrating out the transverse coordinates [e.g. for a 2D channel
$P(x,t) = \int_{-w(x)}^{w(x)} \mathrm{d}y\, P(x,y,t)$]. At variance
with the original Fick-Jacobs equation \cite{Jacobs}, introducing an
$x$-dependent diffusion coefficient considerably improved the
accuracy of the kinetic equation (\ref{eq:fickjacobs}) thus
extending its validity to larger $w(x)$ amplitudes
\cite{Zwanzig,Reguera_PRE,Percus}. Here, the expression
\begin{equation}
  \label{eq:diffusionconst}
  D(x)=\frac{D_{0}}{[1+w'(x)^{2}]^{\alpha}}\, ,
\end{equation}
with $\alpha=1/3$ in 2D and  $\alpha=1/2$ in 3D, was determined to
best account for wall curvature effects
\cite{Reguera_PRE,Burada_PRE}.

In the presence of weak longitudinal drives $F$, the Fick-Jacobs
equation~(\ref{eq:fickjacobs}) can be further extended to
\cite{Reguera_PRE,Burada_PRE,Reguera_PRL}:
\begin{equation}
\label{eq:fickjacobs_ours}
\frac{\partial P}{\partial t}=\frac{\partial}{\partial
  x}D(x)\left(\frac{\partial P}{\partial
    x}+ \frac{A'(x)}{k_{\mathrm{B}}T} P\right)\, ,
\end{equation}
where the free energy $A(x) = E(x) - T S(x)$ is made up of an
energy, $E(x) = - F x$, and an entropic term, $S(x) = k_{\mathrm{B}}
\ln \sigma(x)$. For a periodic channel, $A(x)$ assumes the form of a
tilted periodic potential. Moreover, for a straight channel,
$w'(x)=0$, the entropic contribution vanishes altogether and the
particle is subject to the sole external drive. Of course, for $F=0$
the free energy is purely entropic and
Eq.~(\ref{eq:fickjacobs_ours}) reduces to the Fick-Jacobs equation
(\ref{eq:fickjacobs}). An alternative reduction scheme based on
macrotransport theory has been
proposed recently in Ref.~\cite{Laachi}.

Key quantifiers of the reduced 1D kinetics
(\ref{eq:fickjacobs_ours}) are the average {\it particle current}
or, equivalently, the nonlinear mobility
(\ref{eq:mobility-def}), and the effective diffusion coefficient
(\ref{eq:diffusion-def}) defined in Sec. \ref{sec:exact}. On
generalizing the derivation of Eqs. (\ref{eq:mobility-formula}) and
(\ref{eq:diffusion-formula}) to account for the $x$-dependence of
$D(x)$, one obtains \cite{Reguera_PRL}, respectively,
\begin{equation}
  \label{eq:nonlinearmobility}
  \gamma \mu(f) = \frac{1-e^{-f}}
  {f\,\displaystyle \int_{0}^{L} \frac{\mathrm{d}x}{L} \,I(x)} \, ,
\end{equation}
and
\begin{equation}
  \label{eq:effectivediffusion}
  \frac{D_{\mathrm{eff}}}{D_{0}} = \frac{\displaystyle \int_{0}^{L}
    \frac{\mathrm{d}x}{L} \displaystyle
    \int_{x-L}^{x}\frac{\mathrm{d}z}{L} \,\frac{D(z)}{D(x)}\,
    \frac{e^{A(x)/k_{\mathrm{B}}T}}{e^{A(z)/k_{\mathrm{B}}T}} \,\left[I(z) \right]^2}
  {\left[\displaystyle \int_{0}^{L}\frac{\mathrm{d}x}{L}\,I(x)\right]^3} \, ,
\end{equation}
where
\begin{equation}
\label{eq:Con-I1} I(x) = \frac{e^{A(x)/k_{\mathrm{B}}T}}{D(x)/D_{0}}
\int_{x-L}^{x}\frac{\mathrm{d}y}{L}\,e^{-A(y)/k_{\mathrm{B}}T} \,
\end{equation}
and the dimensionless force $f$ is defined as the ratio of the work
done by $F$ to drag the particle a distance $L$, to the thermal
energy $k_{\mathrm{B}}T$, that is
\begin{equation}
  \label{eq:scalingparameter}
  f = \frac{FL}{k_{\mathrm{B}}T}\, .
\end{equation}

We stress here an important difference with the energetic 1D model
of Sec. \ref{sec:exact}. For a particle moving in a 1D periodic
substrate $V(x)$, the barriers $\Delta V$ separating the potential
wells provide an additional energy scale besides $F L$ and $k_\mathrm{B}T$,
so that the particle dynamics is governed by at least two
dimensionless energy parameters, say, $\Delta V/k_\mathrm{B}T$ and
$FL/k_\mathrm{B}T$ \cite{Risken}. In contrast, Brownian transport in a
periodically corrugated 2D or 3D channel is solely determined by the
dimensionless force $f$ \cite{Reguera_PRL}. This can be proven by rescaling the problem
units as follows. We measure all lengths in units of the period $L$
and time in units of $\tau = L^{2}/D_0$, that is, twice the time the
particle takes to diffuse a distance $L$ at temperature $T$. In
these new dimensionless variables the Langevin equation (\ref{LE})
for a Brownian particle in a static medium reduces to the equation
\begin{equation}
  \label{eq:langevin}
\dot{\vec{x}}(t) = \vec{f} + \vec{\xi}(t)\, ,
\end{equation}
with no tunable constants \cite{Burada_BioSy}. As a consequence, the ensuing Brownian
dynamics is controlled only by the parameter $f$ and, of course, by
the additional length(s) that possibly enter the boundary function
$w(x)$.

The $f$-dependence of the average particle current
(\ref{eq:nonlinearmobility}) and the effective diffusion
(\ref{eq:effectivediffusion}) were compared with the results
obtained by numerical integration of the 2D Langevin equation
(\ref{LE}) (for details see Ref. \cite{Burada_PRE}). For simplicity,
the channel walls were assumed to have the sinusoidal profile
\begin{equation}
w(x) = a [ \sin(2\pi x/L)+ \kappa ]\,,\label{eq:boundary}
\end{equation}
with $a>0$ and $\kappa>1$. Here, $a (\kappa \pm 1)$ are,
respectively, the maximum and the minimum channel width. Moreover,
$a/L$ controls the slope of the boundary function $w(x)$, which in
turn determines the modulation amplitude of the diffusion
coefficient $D(x)$ in Eq. (\ref{eq:diffusionconst}). The mobility
$\mu$, Eq.~(\ref{eq:mobility-def}), and the corresponding effective
diffusion coefficient $D$, Eq.~(\ref{eq:diffusion-num}), were
computed by ensemble averaging $3\cdot 10^{4}$ simulated
trajectories.

\begin{figure}[floatfix]
  \centering
\includegraphics{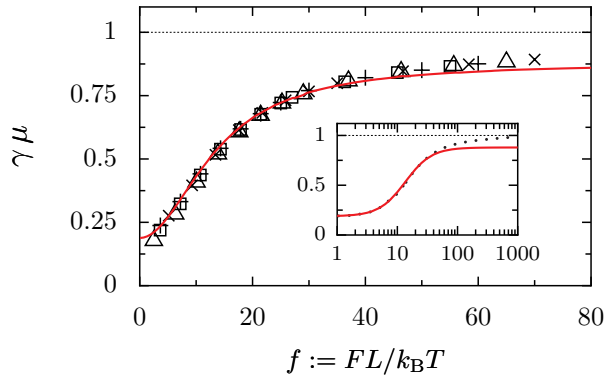}
\caption{(Color online) Nonlinear mobility $\mu$ vs. driving force
in dimensionless units, $f=FL/k_\mathrm{B}T$, for a 2D channel at
different temperatures $k_{\rm B}T=0.01$ (crosses), $0.1$ (pluses), $0.2$ (squares),
and $0.4$ (triangles).
After rescaling, all data sets collapse on one curve which at low $f$
closely compares with the analytic approximation
(\ref{eq:nonlinearmobility}) (solid curve). Other simulation
parameters are $L=1$, $\gamma=1$, and $w(x)=[\sin(2\pi
x)+1.02]/2\pi$. The inset shows the deviation of the analytic
approximation (\ref{eq:nonlinearmobility})  (solid curve) from the
numerical results (dotted curve) for large $f$. Note that the
numerical curve approaches the correct asymptotic limit $\gamma
\mu=1$. }\label{fig:scaledcurrent}
\end{figure}

\begin{figure}[floatfix]
  \centering
\includegraphics{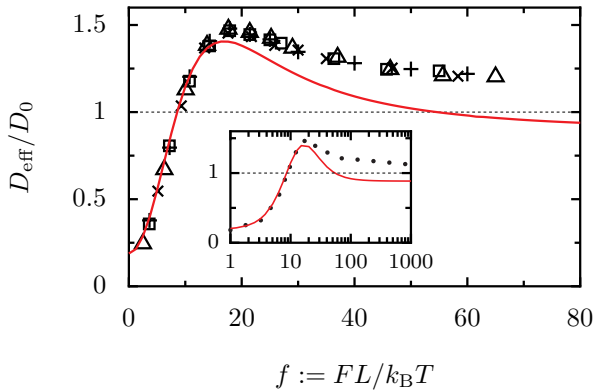}
\caption{(Color online) Effective diffusion coefficient $D_{\rm
eff}$ vs. $f$ for the same simulation parameters as in
Fig.~\ref{fig:scaledcurrent}.  Here, too, the rescaled data collapse
on one curve, which asymptotically approaches the correct limit
$D_{\rm eff}/D_0=1$ (inset). The analytic approximation
(\ref{eq:effectivediffusion}) (solid curve) fits well the raising
branch of the numerical data sets.} \label{fig:scaleddiffusion}
\end{figure}

The most significant results are displayed in
Figs.~\ref{fig:scaledcurrent} and \ref{fig:scaleddiffusion}. At
variance with the purely energetic 1D models of Sec.
\ref{sec:exact}, the nonlinear mobility decreases upon increasing
the strength of thermal noise. Moreover, an enhancement
of the effective diffusion coefficient, with maximum exceeding the
free diffusion constant $D_{0}$, was also observed
\cite{Reguera_PRL,Burada_BioSy}. 

At low values of the control parameter $f$ the analytical
approximations (\ref{eq:nonlinearmobility}) and
(\ref{eq:effectivediffusion}) match perfectly the corresponding
numerical curves, whereas deviations occur at high $f$. Most
remarkably, contrary to the simulation output, the analytical curves
for $D_{\rm eff}/D_0$ and $\gamma \mu$ fail to attain the correct
asymptotic limit $1$ for $f\to\infty$ (see insets in
Figs.~\ref{fig:scaledcurrent} and \ref{fig:scaleddiffusion}). This
occurs because the assumption of uniform density distribution,
introduced in the Fick-Jacobs formalism to eliminate the transverse
coordinates, is no more tenable in the presence of strong drives.

The agreement between theory and numerics improves for smooth
modulations of the channel walls, i.e. for small boundary slopes
$|w^{\prime}(x)|$, which can be achieved, e.g., for appropriately
small $a$ \cite{Burada_PRE,Burada_BioSy}. A phenomenological
criterion to asses the validity of the stationary state solutions of
the Fick-Jacobs equation (\ref{eq:fickjacobs_ours}) can be
formulated by comparing the different characteristic time scales of
the problem, namely \\ (i) the diffusion times, respectively, in the
transverse,
\begin{equation}
\tau^{(d)}_{\perp} = (2 a^{2}/D_0) (1+\kappa)^{2} \label{tauT},
\end{equation}
(ii) and in the longitudinal direction,
\begin{equation}
\tau^{(d)}_{\|} = L^{2}/2D_0,
\end{equation}
(iii) and the drift time the applied force $F$ takes to drag the
particle one channel unit length $L$ across,
\begin{equation}
\tau^{(f)}_{\|} = \gamma L/F \, . \label{taudf}
\end{equation}

Uniform probability distribution in the transverse direction sets on
only if the transverse diffusion motion of the particle is
sufficiently fast relative to both the diffusive and the drift
longitudinal motions, which implies $\tau^{(d)}_{\perp} \ll {\rm
 min}\{\tau^{(d)}_{\|},\tau^{(f)}_{\|}\}$.
For large drives it suffices to require that $\tau^{(d)}_{\perp} \ll
\tau^{(f)}_{\|}$, which leads to the condition \cite{Burada_BioSy}:

\begin{equation}
f \leq f_c \equiv\frac{1}{2(1+\kappa)^{2}} \,
\left (\frac{L}{a}\right )^2\, . \label{c2}
\end{equation}
\begin{figure}[floatfix]
  \centering
  \includegraphics{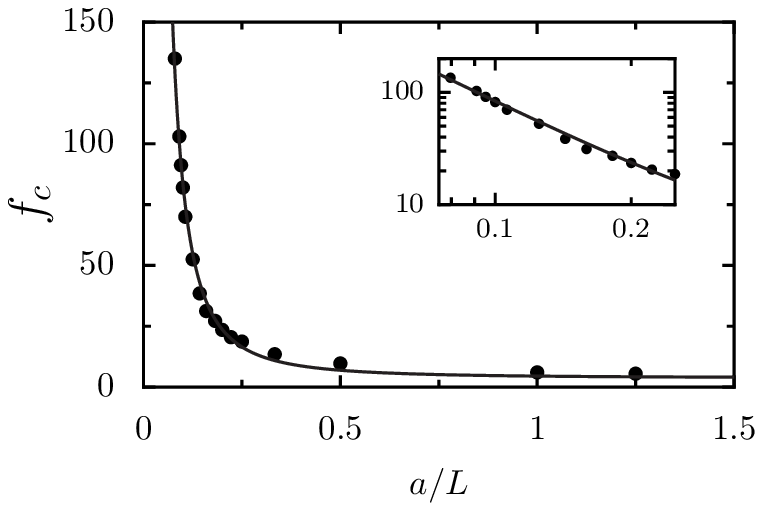}
  \caption{Critical force $f_c$ vs. boundary modulation amplitude 
    $a$ for the 2D channel
    of Fig. \ref{fig:scaledcurrent}. Other simulation
    parameters are $L=1$, $\gamma=1$, and $k_{\rm B}T=0.1$. The
    numerical values of $f_c$ (solid dots) were obtained by requiring
    deviations of Fick-Jacobs approximation from the simulation results
    smaller than $\sim 1\%$. The solid line is the fitting law
    $f_{c}=c_{1} (L/a)^{2} + c_{2}$ with $c_{1}=0.792$ and $c_{2}=3.524$. 
    The apparent discrepancy with Eq.~(\ref{c2}) results
    from the above-mentioned numerical accuracy criterion and is
    expected to 
    vanish for higher accuracy. Inset: same data sets on a logarithmic
    scale.}
  \label{fig:critical}
\end{figure}

The critical force parameter $f_c$, above which the Fick-Jacobs
description is expected to fail, depends on the remaining free
parameters of the problem. For the boundary function
(\ref{eq:boundary}), $f_c$ is a function of $a/L$, alone, as
illustrated in Fig.~\ref{fig:critical}. For $a\gg L$ the Fick-Jacobs
approximation becomes untenable already for relatively small forces
$f$, whereas for $a\ll L$ its validity extends to significantly
larger drives.

\section{Particle diffusion in moving fluids}
\label{sec:4.2}

Let us assume now that a small, spherical particle is swept along in
the stationary velocity field $\vec{v}(\vec{x},t)$ of a moving,
incompressible fluid [see Fig.~\ref{fig:tube}(b)], rather than by a constant force,
$\vec{F}_{\mathrm{ext}}$, like in Sec. \ref{sec:4.1}.  For
$\vec{F}_{\mathrm{ext}}=0$, its dynamics is then described by the
Langevin equation, in the overdamped limit,
\begin{equation}
  \label{LEMF}
  \dot{\vec{x}}({t}) = \vec{v}(\vec{x}({t}),{t})
  +\sqrt{\frac{2 k_\mathrm{B}T}{\gamma}} \vec{\xi}({t})\, ,
\end{equation}
where the fluid velocity field $\vec{v}(\vec{x},t)$ is to be
determined.

\begin{figure}[t]
  \centering
 \label{fig:particleseparation}
\includegraphics[width=7.5truecm]{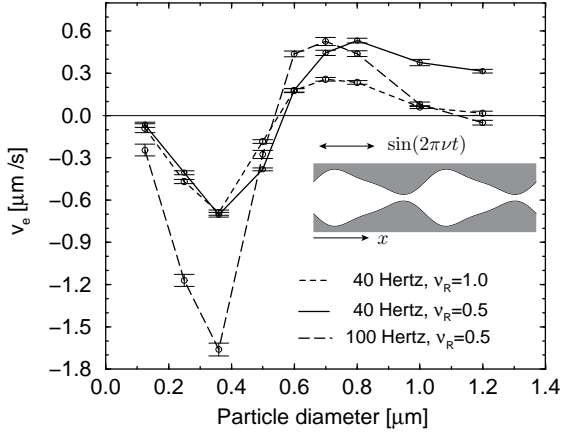}
\caption{Average induced particle current $\mathrm{v}_\mathrm{e}$  versus particle
diameter. The particle is immersed in a laminar flow confined to a
channel with broken reflection symmetry (inset). 
The fluid, with viscosity
$\nu_\mathrm{R} \equiv \eta / \eta_{\mathrm{water}}$ and
$\eta_{\mathrm{water}}=1.025\cdot 10^{3} \mathrm{Ns / m}^{2}$, is
pumped periodically back and forth, along the $x$ axis of the channel,
with frequency $\nu$. 
For further details, see Ref. \cite{kettner}.}
\end{figure}

To this purpose, we write first the Stokes equations  for a
stationary incompressible flow in the limit of vanishing Reynolds
numbers (the convective acceleration terms in the Navier-Stokes
equations can be safely neglected). One obtains the so-called
``creeping flow equation'' \cite{Lamb}: $\eta \Delta
\vec{v}(\vec{x}) = \vec{\nabla} p( \vec{x} )\, ,$ where
$\vec{p}(\vec{x})$ is the pressure field responsible for the
stationary laminar flow of the fluid and $\eta$ is the absolute
dynamic fluid (or shear) viscosity. On introducing the scalar field
$\Psi(x,r)$, with $(r,\phi)$ being the polar coordinates in the
transverse plane $(y,z)$, $\vec{v}(\vec{x})$ can be rewritten as
\cite{Lamb}
\begin{equation}
  \label{eq:fv}
  \vec{v}(x, r) = \vec{\nabla} \times \left( \Psi(x, r) \vec{e}_{\phi}/r\right)\,
  ,
\end{equation}
where $\vec{e}_{\phi}$ is the unit $\phi$  vector. On substituting
Eq.~(\ref{eq:fv}) into the creeping flow equation, one obtains the
following linear homogeneous fourth-order differential equation
\cite{Lamb, kettner}:
\begin{equation}
  \label{eq:streamlinefunction}
  \left( r \frac{\partial }{\partial r} \frac{1}{r} \frac{\partial
    }{\partial r} + \frac{\partial^{2}}{\partial x^{2}}\right)^{2}
  \Psi(x, r) = 0\, ,
\end{equation}
with boundary conditions
\begin{eqnarray}
  \label{eq:slfbca}
  &&\Psi|_{r=0} = c \, ,\\
  \label{eq:slfbcb}
  &&\frac{\partial}{\partial r}\Psi|_{r=0} =\frac{\partial^3}{\partial r^3}\Psi|_{r=0}=0\, ,\\
  \label{eq:slfbcc}
  &&\vec{\nabla} \Psi(x,r=w(x))=0\, ,\\
  \label{eq:slfbcd}
  &&\Psi(x+L,r) = \Psi(x, r)\, ,
\end{eqnarray}
where $c$ is an arbitrary constant.
Solutions to Eq.~(\ref{eq:streamlinefunction}), 
determine the velocity field  $\vec{v}(\vec{x})$ up to a
multiplicative factor, which, in turn, can be established by
imposing some additional condition, e.g. for the pressure drop
across a channel unit, i.e. \cite{kettner}
\begin{equation}
  \label{eq:multconst}
  p(x,r)-p(x+L,r) = -2 \int_{0}^{L} \left (\frac{\partial^{2}}{\partial
    r^{2}} v_{x}\right )_{r=0} \mathrm{d} x\, .
\end{equation}
Once the velocity field for a particular channel geometry and
pressure profile is known, the Langevin equation (\ref{LEMF}) can be
solved numerically.

Particle separation across a micro-channel is a process of great
technological importance \cite{BM1,BM}. Any inhomogeneity in the
spatial distribution of an ensemble of non-interacting suspended
particles can only be caused by the hydrodynamic interaction between
particles and walls \cite{Schind}. The no-flux boundary condition
(\ref{bcw}) for unbiased particles with
$\vec{F}_{\mathrm{ext}}=\vec{0}$ reads:
\begin{equation}
  \label{eq:bcparticles}
  \vec{n}(\vec{x}) \cdot \left( \vec{v}(\vec{x}) P(\vec{x},t) -
    \frac{k_{\mathrm{B}}T}{\gamma} \vec{\nabla} P(\vec{x},t)
  \right) = 0\, , \quad \vec{x} \in \mathrm{ wall}\, .
\end{equation}
In the case of vanishing drift velocity normal to the channel walls,
$\vec{n}(\vec{x}) \cdot \vec{v}(\vec{x}) = 0$, only  uniform
distributions are allowed and no particle separation could ever be
achieved \cite{Schind}. However, as anticipated in Sec.
\ref{sec:bc}, for finite size particles the no-flux boundary
condition (\ref{eq:bcparticles}) strictly holds on an effective
inner surface at a distance from the channel walls. Due to its
finite size, a particle cannot move steadily along a given flow
streamline as this gets too close to the walls; upon hitting the
wall, the particle bounces into inner flow streamlines as if they
were subject to a transverse field gradient. These hydrodynamic
forces may, indeed, lead to accumulation and depletion zones within
the channel.

By generalizing such a boundary effect, Kettner {\itshape et al.}
\cite{kettner} predicted by numerical simulation that a micron-sized
channel with broken reflection symmetry can be used to separate
particles according to their size, as illustrated in
Fig.~\ref{fig:particleseparation}. A time oscillating pressure
profile, $\vec{p}(\vec{x},t) = \vec{p}(\vec{x}) f(t)$, where $f(t)$
is a sinusoidal function with frequency $\nu$, was assumed to
control the periodic flow of the fluid, back and forth along an
infinitely long channel. Within the creeping flow approximation, the
ensuing time dependent velocity field $\vec{v}(\vec{x},t) =
\vec{v}(\vec{x}) f(t)$ was then obtained in terms of the solution
$\vec{v}(\vec{x})$ of the {\itshape unperturbed} Stokes equation
introduced above, and the corresponding Langevin equation
(\ref{LEMF}) numerically integrated. Later on, this mechanism was
demonstrated experimentally by Matthias and M{\"u}ller
\cite{muller}.


\section{Single File Diffusion}
\label{sec:singlefile}

As the design and the operation of biology inspired nanodevices
\cite{BM} have become experimentally affordable, understanding
particle diffusion in a 1D substrate has been recognized as a key
issue in transport control \cite{machura}. In this context the
particle- and particle-wall interactions play a central role. Pair
interaction between thermally diffusing particles does not affect
the normal character of Brownian diffusion, as long as the particles
are able to pass one another, no matter how closely they are
confined. This holds true even when, under appropriate temperature
and density conditions, attracting particles cluster or condense in
the substrate wells \cite{our_prl}.

Things change dramatically for strictly 1D geometries. Let us
consider, for instance, an ensemble of $N$ unit-mass particles
moving with preassigned dynamics along a segment of length $l$. If
the interparticle interaction is hard-core (with zero radius), the
elastic collisions between neighboring particles are nonpassing --
meaning that the particles can be labeled according to a fixed
ordered sequence. The particles are thus arranged into a file where,
at variance with the situation described in Sec. \ref{sec:model},
their diffusion is suppressed by the presence of two nonpassing
neighbors, also movable in the longitudinal direction.

\begin{figure}[t]
  \centering
\includegraphics*[width=8.0cm,clip]{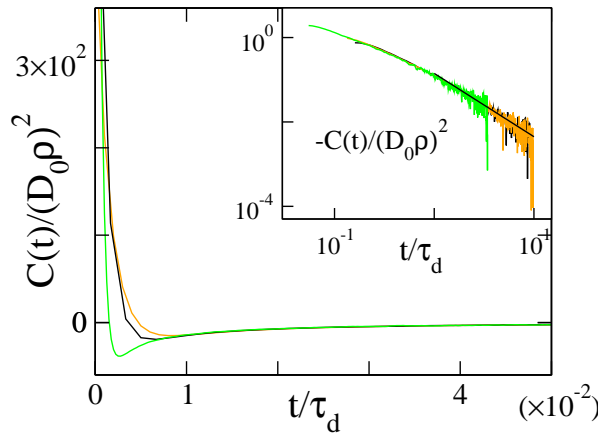}
\caption{\label{fig:6.1} (Color online) Velocity autocorrelation
function $C(t)$ in a sSF at $k_{\rm B}T=1$ and for $(\rho,\gamma)=(0.1,3)$
(top, orange curve), $(0.1,5)$ (middle, black curve), and $(0.05,5)$
(bottom, green curve). Inset: log-log plot of the negative $C(t)$
tails; the solid line is the predicted tail $-C(t)/(D_0\rho)^2
\simeq (t/\tau_d)^{-\frac{3}{2}}/(4\sqrt{\pi})$.}
\end{figure}

The diffusion of a free single file (SF), i.e. in the absence of a
substrate, has been investigated in detail \cite{SFD-theory}. In the
thermodynamic limit ($l,N \to \infty$ with constant density
$\rho\equiv N/l$) the mean square displacement of each file particle
can be written as
\begin{equation}
\label{eq:6.1} \langle \Delta x^2(t) \rangle = |\Delta x(t)|/\rho
\end{equation}
with $|\Delta x(t)|$ denoting the absolute mean displacement of a
free particle. For a {\it ballistic} single file (bSF), clearly
$|\Delta x(t)|=\langle |v|\rangle t$, where $\langle \dots \rangle$
is the ensemble average taken over the distribution of the initial
velocities, and therefore
\begin{equation}
\label{eq:6.2} \langle \Delta x^2(t) \rangle = \langle |v| \rangle
t/\rho.
\end{equation}

A bSF particle diffuses apparently like a Brownian particle with
normal diffusion coefficient $D_0=\langle |v| \rangle/(2\rho)$. For
a {\it stochastic} single file (sSF) of Brownian particles with
damping constant $\gamma$ at temperature $T$, the equality $|\Delta
x(t)|=\sqrt{4D_0t/\pi}$ yields the {\it anomalous} diffusion law
\begin{equation}
\label{eq:6.3} \langle \Delta x^2(t) \rangle = 2D_{\rm
SF}\sqrt{t}/\rho,
\end{equation}
where the mobility factor $D_{\rm SF}=\sqrt{D_0/\pi}$ is related to
the single particle diffusion constant $D_0=k_{\rm B}T/\gamma$.  The onset
of the subdiffusive regime (\ref{eq:6.3}) occurs for $t > \tau_d$,
$\tau_d=(D_0\rho^2)^{-1}$ being the average time a single particle
takes to diffuse against one of its neighbors \cite{hop,taloni2}.
The diffusive regimes (\ref{eq:6.2}) and (\ref{eq:6.3}) have been
demonstrated both numerically \cite{6simulations} and experimentally
\cite{6bechinger,6experimental}.
\begin{figure}[t]
  \centering
\includegraphics*[width=7.0cm,clip]{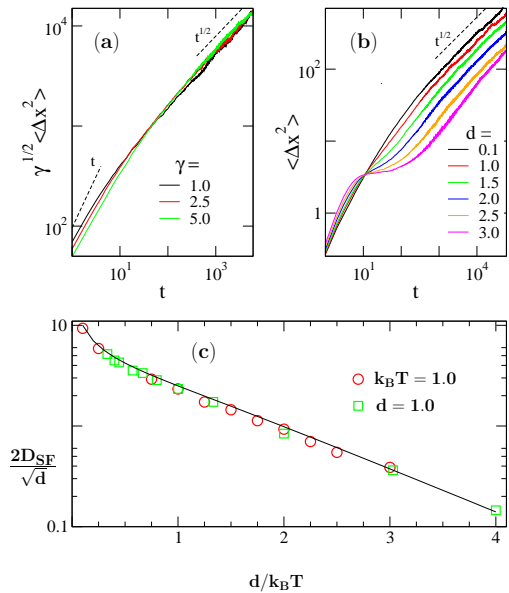}
\caption{\label{fig:6.2} (Color online) Diffusion of a stochastic
single file in the periodic potential (\ref{eq:6.5}) \cite{taloni1}:
(a) $\langle \Delta x^2(t) \rangle$ vs. $t$ for $k_{\rm B}T=1$, $d=1$ and
increasing $\gamma$ (from top to bottom on the l.h.s); (b) $\langle
\Delta x^2(t) \rangle$ vs. $t$ for $\gamma=5$ and increasing $d$
(from top to bottom on the r.h.s). The $t$ and $t^{1/2}$ slopes
(dashed lines) have been drawn for reader's convenience; (c) the
diffusion mobility $D_{\rm SF}$ vs. $d/k_{\rm B}T$ for $\gamma=5$ and
$d=1$ (circles) and $k_{\rm B}T=1$ (squares). Other simulation parameters
are: $N=3\times 10^3, L=2\pi, l=3 \times 10^3 L$, and all particles
have unit mass.}
\end{figure}
Let $x(t)$ represent the coordinate of one file particle, assumed to
be a continuous differentiable stochastic process with $\langle \dot
x(t) \rangle =0$. Kubo's relation \cite{kubo}
\begin{equation}
\label{eq:6.4} \frac {1}{2}\,\frac{d}{dt}\, \langle\Delta^2
x(t)\rangle = \int _0^t C(\tau)d\tau ~~~~(t\to \infty),
\end{equation}
with $C(t)\equiv \langle \dot x(t)\dot x(0) \rangle$, best
illustrates the role of the dimensional constraint on SF diffusion.
In the case of normal diffusion, $\langle\Delta^2x(t)\rangle = 2D_0
t$, the r.h.s. of Eq. (\ref{eq:6.4}) converges to the positive value
$D_0=\int_0^{\infty}C(\tau)d\tau$, namely $x(t)$ diffuses subject to
Einstein's law. The subdiffusive dynamics (\ref{eq:6.3}), instead,
is characterized by $\int_0^{\infty}C(\tau)d\tau=0$, which, in view
of Eq. (\ref{eq:6.4}), implies that $C(t)$ develops a negative
power-law tail $C(t) \sim -c_\beta t^{-\beta}$ with
$\beta=\frac{3}{2}$ and $c_\beta=D_{\rm SF}/4\rho$. Numerical
simulations support this conclusion. In Fig. \ref{fig:6.1} we report
a few curves $C(t)$ for different $\rho$ and $\gamma$. The negative
tails are apparent and compare well with the estimate derived from
Eq. (\ref{eq:6.4}) (see inset). They are a typical signature of the
collisional dynamics in a SF: As an effect of pair collisions, an
initial velocity $\dot x(0)$ is likely to be compensated by a
backflow velocity of opposite sign \cite{taloni2,taloni3}.

\subsection{Stochastic single file on a substrate}
\label{sSF}

We focus now on the case of a SF diffusing on a sinusoidal substrate
with potential \cite{percus2,taloni1}
\begin{equation}
\label{eq:6.5} V(x)=d[1 - \cos (2\pi x/L)].
\end{equation}
This variation of the SF model rises naturally in connection with
quasi-1D situations where the particles (not necessarily suspended
in a fluid) can be represented by disks moving along a narrow
spatially modulated channel with cross-section smaller than twice
the disk diameter [see Fig. \ref{fig:tube}(c)]. The confining action
of the channel can be accounted for by modeling the SF dynamics in
terms of a periodic substrate potential with a effective strength
$d$. This is the case, for instance, of most nanotubes and zeolite
pores \cite{zeoliteK}.

In the simulations of Refs. \cite{taloni2,taloni1} the $i$-th
particle was assigned: (i) random initial position, $x_i(0)$, and
velocity, $\dot x_i(0)$. Upon each elastic collision it switched
velocity with either neighbors without altering the file labeling;
(ii) independent Brownian dynamics determined by a viscous force
$-\gamma \dot x_i$ and a thermal force $\sqrt{2\gamma k_{\rm
B}T}\xi_i(t)$, with $\xi_i(t)$ uncorrelated standard Gaussian noises
defined as in Sec. \ref{sec:diffequations}, in order to guarantee
thermalization at temperature $T$.

Numerical evidence led to conclude that the periodic substrate
potential $V(x)$ does not invalidate the sSF diffusion law
(\ref{eq:6.3}), although the dependence of the mobility factor
$D_{\rm SF}$ on the system parameters becomes more complicated. In
panels (a) and (b) of Fig. \ref{fig:6.2} $D_{\rm SF}$ is
characterized as a function of $\gamma$ and $T$, respectively. The
identity $D_{\rm SF}=\sqrt{D_0/\gamma}$, reported above for
$V(x)\equiv 0$, applies here, too, under the condition of replacing
$D_0$ with the modified diffusion constant $D(F)$ of Eq.
(\ref{eq:diffusion-formula}). In Fig. \ref{fig:6.2}(a) the rescaled
curves $\gamma^{1/2}\langle \Delta x^2(t) \rangle$ versus $t$
overlap asymptotically for any damping regime.

The temperature dependence of the mobility $D_{\rm
SF}=\sqrt{D(F)/\gamma}$, is more interesting. As implicit in the
discussion of Sec. \ref{sec:exact}, this prediction gets more and
more accurate for large $\gamma$ and increasingly high
activation-to-thermal energy ratios $d/k_{\rm B}T$. As a consequence, the
rescaled mobility $\sqrt{\gamma/d}~D_{\rm SF}$ turns out to be a
function of $d/k_{\rm B}T$, alone, in good agreement with the simulation
results displayed in Fig. \ref{fig:6.2}(c).
\begin{figure}[ht]
  \centering
\includegraphics*[width=7.0cm]{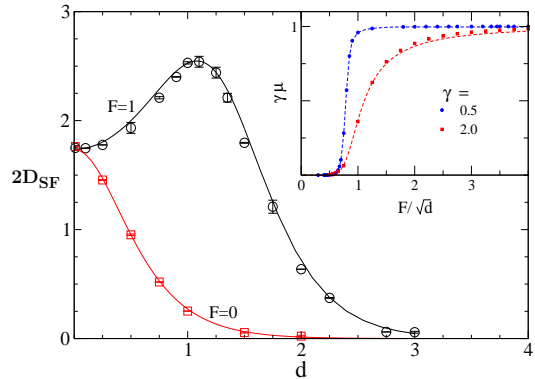}
\caption{\label{fig:6.3} (Color online) Diffusion of a driven
stochastic single file in the periodic potential (\ref{eq:6.5}): the
mobility factor $D_{\rm SF}$ vs $d$ for $k_{\rm B}T=0.3$, $\gamma=5$, and
$F=0$ (squares) and $F=1$ (circles). The solid curves represent the
law $D_{\rm SF}=\sqrt{D(F)/\pi}$ with $D(F)$ given in Eq.
(\ref{eq:diffusion-formula}). Other simulation details are as in
Fig. \ref{fig:6.2}. Inset: the kinetic mobility, $\mu = \langle \dot
x \rangle/F$ vs. $F/\sqrt{d}$ for $\gamma=0.5$ (circles) and
$\gamma=2$ (squares) at $k_{\rm B}T=0.1$. The fitting curves are the
analytical predictions for a single Brownian particle based,
respectively, on Eqs. (11.194) (low $\gamma$) and (11.50) (high
$\gamma$) of Ref. \cite{Risken}.}
\end{figure}

\subsection{Driven single files}
\label{driven SF}

Finally, we consider the case of a {\it driven} sSF, namely, we now
assume that all file particles are subjected to an additional
constant force $F$ pointing, say, to the right ($F \geq 0$).  We
know from Sec. \ref{sec:exact} that the diffusion of a single
Brownian particle drifting down a tilted washboard potential
exhibits enhanced normal diffusion with diffusion constant
(\ref{eq:diffusion-formula}). Extensive simulation of a driven SF
\cite{taloni1} yielded the numerical data reported in Fig.
\ref{fig:6.3}. The kinetic mobility of the file, defined as $\mu =
\langle \dot x \rangle/F$, turned out to coincide with the mobility
of a single particle under the same dynamical conditions,
irrespective of $\gamma$ (inset of Fig. \ref{fig:6.3}). More
remarkably, the subdiffusive regime (\ref{eq:6.3}) applies also in
the presence of bias, though with an $F$-dependent mobility factor.
When plotted versus $d$, $D_{\rm SF}$ attained a maximum enhancement
for $d \geq F$, i.e., in coincidence with the (noise-assisted)
depinning of the file from its sinusoidal substrate. Again, the
identity $D_{\rm SF} =\sqrt{D(F)/\pi}$ combined with formula
(\ref{eq:diffusion-formula}) for $D_0$ provided an excellent fit of
the simulation data for large $\gamma$. Of course, the mobility
enhancement at depinning can be revealed also by plotting $D_{\rm
SF}$ versus $F$ at constant $d$.

\section{Conclusions and Outlook}

With this minireview we presented the state of the art of diffusive
transport occurring in systems characterized by confinement due to
either their finite size or to particle interactions in restricted
geometries. Confinement plays a salient role in the Brownian motion
of driven particles. Indeed, the entropic effects associated with
confinement may give rise to anomalous transport features. Some main
new phenomena are the entropy-driven decrease of mobility with
increasing temperature, the diffusion excess above the free
diffusion limit and the resonant behavior of the effective diffusion
coefficient as a function of control parameters such as temperature,
external gradients, or geometry design.

The reviewed transport features can thus be implemented in the
design of new transport control setups or protocols. In particular,
Brownian transport through confined geometries is expected to find
application in nanotechnology, for instance, to operate devices for
the separation of nanoparticles, to speed up schemes for catalysis
on templates, or also to realize efficient, driven through-flows of
microsized agents, or reactants, in miniaturized lab-on-a-chip
devices.

\section*{Acknowledgements}
This work was made possible thanks to the financial support by: the
Volkswagen Foundation (project I/80424, P.H.); the Deutsche
Forschungsgemeinschaft (DFG) via project no. 1517/26-1 (P.S.B.,
P.H.) and via the research center, SFB-486, project A10 (G.S.,
P.H.); the German Excellence Initiative, via the \textit
{Nanosystems Initiative Munich} (NIM) (P.S.B., P.H.); and the
Alexander von Humboldt Stiftung, via a Research Award (F.M.).


\end{document}